\documentclass[conference]{IEEEtran}
\IEEEoverridecommandlockouts

\usepackage[numbers,sort&compress,sectionbib]{natbib} 
\usepackage{amsmath,amssymb,amsfonts}
\usepackage{algorithm}
\usepackage{algorithmic}
\usepackage{graphicx}
\usepackage{textcomp}
\usepackage{xcolor}
\usepackage{booktabs}
\usepackage{url}
\usepackage{array}
\usepackage[table]{xcolor}
\usepackage{float}

\newcommand{\sailer}[1]{\cellcolor{yellow!20}{#1}}

\usepackage{tabularx}
\usepackage{ragged2e}
\usepackage{tabularx}
\newcolumntype{L}{X}
\newcolumntype{C}{>{\centering\arraybackslash}X}
\newcolumntype{R}{>{\raggedleft\arraybackslash}X}
\usepackage{multirow}

\usepackage{lipsum}

\usepackage{hyperref}
\hypersetup{
colorlinks   = true,
citecolor    = blue,
urlcolor    =  blue
}

\begin{document}

\title{Section-Weighted Hybrid Approach for Legal Case Retrieval}

\author{
\IEEEauthorblockN{Rajith Arulanandam, Nisansa de Silva\thanks{*Corresponding author: Nisansa de Silva (\texttt{nisansadds@cse.mrt.ac.lk}).}}
\IEEEauthorblockA{Department of Computer Science \& Engineering, University of Moratuwa, Sri Lanka\\
\texttt{\{rajith.24,nisansadds\}@cse.mrt.ac.lk}}
}

\maketitle

\begin{abstract}
Finding truly analogous precedents requires capturing legal reasoning beyond surface word overlap. We present a two-stage, \emph{section-aware} framework for legal case retrieval that first segments raw judgments into \textit{facts}, \textit{issues}, \textit{decision}, and \textit{reasoning} using a deterministic large language model (LLM) offline. In \textbf{Stage~1}, we combine parallel lexical (BM25) and semantic (dense ANN) whole-document searches via Reciprocal Rank Fusion (RRF) to form a high-recall candidate pool. In \textbf{Stage~2}, we perform fine-grained, like-for-like comparisons (e.g., query reasoning vs. candidate reasoning). To address the scale mismatch between unbounded lexical scores and cosine similarities, we apply query-wise Z-score normalization before aggregating signals with learned section weights. For the top results, the system returns the relevant section text with a concise, grounded rationale and \emph{party-stance} labels. We evaluate on a jurisdiction-scale benchmark, demonstrating consistent gains over strong lexical and neural baselines while maintaining high candidate coverage.
\end{abstract}


\section{Introduction}
Legal research demands finding precedents that match in \emph{reasoning}, not just text~\cite{sugathadasa2018legal}. Classical methods such as BM25~\cite{article} struggle when key arguments are phrased differently, while modern single-vector models~\cite{jayasinghe2022learning,jayawardana2017word} often blur legally distinct facets such as \textit{facts} and \textit{decisions}~\cite{ratnayaka2022context}. This inability to distinguish between what happened and why it matters creates a precision--recall trade-off that hampers reliable discovery.

Recent advances have addressed parts of this gap. Some systems refine queries or leverage case-based reasoning to find analogies~\cite{deng2024keller,wiratunga2024cbr}, while others focus on explainability and snippet-level evaluation~\cite{yu2022explainable,pipitone2024legalbench}. Domain-specific encoders~\cite{jayawardena2024scale,licari2024italian} and graph-based models~\cite{bhattacharya2020hier,sun2023law} further demonstrate the value of jurisdiction and citation signals. Additionally, logic-informed and judge-style LLM evaluations highlight the growing importance of interpretability~\cite{sun2024logic,ge2021learning,wu2023precedent,zheng2023judging,huang2024empirical,chen2024humans}.

To unify these strengths, we propose a transparent, two-stage \emph{section-aware} architecture. Before retrieval, an offline LLM organizes judgments into four facets facts, issues, decision, and reasoning storing them as structured text fields. \textbf{Stage~1} casts a wide net by fusing keyword and semantic searches to capture all potential candidates. \textbf{Stage~2} then performs a precise, \emph{like-for-like} comparison (e.g., query reasoning vs. candidate reasoning), aggregating scores with learned weights. Finally, the system returns the relevant section text and a concise rationale with \emph{party-stance} labels.

Our contributions include: (i) a scalable retrieval framework with explicit section-aware scoring; (ii) an offline facetization method enabling structured, grounded evidence; and (iii) experiments demonstrating consistent gains over strong baselines on large-scale benchmarks.

\section{Related Work}

\subsection{Jurisdiction-Tailored Embeddings}
Tailoring embeddings~\cite{jayasinghe2022learning,jayawardana2017word} to jurisdiction-specific corpora~\cite{sugathadasa2017synergistic} improves retrieval effectiveness~\cite{sugathadasa2018legal,jayasinghe2022legal}. In the context of low-resource legal domains, \citet{jayawardena2024scale} utilized BM25-ranked triplets to fine-tune AnglE-BERT via contrastive learning. Their work demonstrates that generating dual embedding representations specifically optimized for either symmetric semantic similarity or asymmetric information retrieval significantly enhances retrieval F1-scores compared to standard baselines. Similarly, Italian-Legal-BERT~\cite{licari2024italian} showed comparable gains for the Italian legal system. 

Beyond text-only modeling, incorporating graph signals such as citation and statute linkages further enhances retrieval. \citet{bhattacharya2020hier} proposed Hier-SPCNet, a heterogeneous network that augments standard citation graphs with the hierarchical structure of legal statutes (e.g., Acts, Chapters, Sections). By applying metapath-based graph embeddings, they demonstrated that explicitly modeling these statutory relationships provides complementary signals to text-based retrieval methods. Furthermore, Law-Match~\cite{sun2023law} introduces a causal learning framework that utilizes law articles as instrumental variables to decompose legal case embeddings into ``legal essence'' and ``factual noise.'' By filtering out irrelevant details through this mediation analysis, the model significantly improves retrieval accuracy in scenarios where semantic similarity does not align with legal relevance.

\subsection{Explainability and Legal Logic Modeling}
Explainability is a core requirement for trustworthy legal AI. NS-LCR~\cite{sun2024logic} combines symbolic logic rules with neural retrieval to align reasoning steps with case outcomes, improving interpretability. \citet{ge2021learning} directly align fact patterns with statutory provisions to produce structured justifications. These approaches promote verifiable chains of reasoning that complement score-only retrieval outputs.

\subsection{Case-Based Reasoning in Legal RAG}
Case-based reasoning (CBR) mirrors practitioner workflows by retrieving and adapting analogical precedents~\cite{jayasinghe2022legal}. CBR-RAG conditions retrieval on similarity across intra-case and inter-case structures, yielding better factual grounding and citation usage than retrieval-agnostic generation on ALQA~\cite{wiratunga2024cbr}.

\subsection{LLM-Enhanced Retrieval and Query Reformulation}
Large language models (LLMs) have improved zero-shot performance on legal tasks. Providing precedents to LLMs boosts judgment prediction quality in PLJP~\cite{wu2023precedent}. KELLER~\cite{deng2024keller} refines queries by decomposing them into statute-anchored sub-facts using knowledge-guided prompting, improving interpretability and precision via contrastive learning over aligned elements.

\subsection{Evaluation at the Snippet Level}
Document-level recall can mask failures to retrieve the exact evidentiary text needed in practice~\cite{sugathadasa2018legal}. LegalBench-RAG introduces snippet-level supervision with 6{,}858 QA pairs annotated at minimal spans, reframing evaluation toward precise evidence retrieval and better matching how lawyers search and cite~\cite{pipitone2024legalbench}.

\subsection{LLM-as-a-Judge: Capabilities and Limitations}
LLM judges have shown promising agreement with humans~\cite{gunathilaka2025automatic}: MT-Bench and Chatbot Arena report over 80\% agreement for GPT-4, comparable to inter-human consistency~\cite{zheng2023judging}. However, subsequent studies warn of overfitting and limited generalization when fine-tuning judge models~\cite{huang2024empirical}, as well as persistent biases (e.g., authority and gender bias) in both human and LLM evaluations~\cite{chen2024humans}. These findings underscore the need for transparent, bias-aware designs and grounded outputs in high-stakes legal settings.

\section{Methodology}

\subsection{Offline Pre-processing: Facet Extraction}
\textbf{Input:} Raw full-text judgments. \textbf{Output:} Four structured facet texts ($d_{\mathrm{facts}}, d_{\mathrm{issues}}, \dots$).
Each judgment is processed by a deterministic LLM (temperature $=0$) to extract facets. We persist the extracted text for each section, enabling direct retrieval of specific legal arguments without re-processing. We also compute and store embeddings for both the whole document ($E(d_{\mathrm{whole}})$) and individual sections ($E(d_i)$).

\subsection{Stage 1: Candidate Generation (Fast, Coarse)}
We build a BM25 index over $d_{\mathrm{whole}}$ and a dense index over $E(d_{\mathrm{whole}}) \in \mathbb{R}^D$. Given query $q$ with $q_{\mathrm{whole}}$ and embedding $E(q_{\mathrm{whole}})$, we retrieve top-$A$ BM25 and top-$B$ dense sets, then fuse by RRF:
\begin{align}
\mathrm{RRF}(d) = \frac{1}{k + r^{\mathrm{BM25}}(d)} + \frac{1}{k + r^{\mathrm{Dense}}(d)}, \quad k=60,
\end{align}
with missing ranks contributing $0$. The candidate pool $\mathcal{P}(q)$ is the top-$M$ by RRF.

\subsection{Stage 2: Section-Aware Re-ranking}
For each candidate $d$ in the pool $\mathcal{P}(q)$, we compute similarities for every section $i \in \{\text{facts}, \text{issues}, \text{decision}, \text{reasoning}\}$.

\subsubsection{Score Computation}
For a query section $q_i$ and document section $d_i$:
\begin{align}
    s^{\text{lex}}_{i} &= \mathrm{BM25}(q_i, d_i) \\
    s^{\text{sem}}_{i} &= \cos(E(q_i), E(d_i))
\end{align}

\subsubsection{Aggregation and Normalization}
We aggregate section scores using learned weight vectors $\mathbf{w}^{\text{lex}}$ and $\mathbf{w}^{\text{sem}}$. Since BM25 scores are unbounded and cosine similarity is $\in [-1, 1]$, direct summation is unstable. We apply \emph{per-query Z-score normalization} ($\mathcal{Z}$) to the aggregated signals:
\begin{align}
    S_{\text{lex}}(q,d) &= \mathcal{Z}\left( \sum_{i} w^{\text{lex}}_i \cdot s^{\text{lex}}_{i} \right) \\
    S_{\text{sem}}(q,d) &= \mathcal{Z}\left( \sum_{i} w^{\text{sem}}_i \cdot s^{\text{sem}}_{i} \right)
\end{align}
The final ranking score is a weighted fusion of these normalized signals:
\begin{equation}
    \mathrm{Score}(q,d) = \alpha \cdot S_{\text{lex}}(q,d) + \beta \cdot S_{\text{sem}}(q,d)
\end{equation}
where $\alpha$ and $\beta$ are hyperparameters tuned on the development set.
\subsection{Explanation, Grounding, and Party Stance}
For the top-$K$ results, we retrieve the relevant section text (e.g., the specific reasoning block that triggered the match) and call a deterministic LLM (temperature $=0$) for a 1--3 sentence rationale and \emph{party-stance} labels (supports/neutral/opposes).

\subsection{Hyperparameters (Default)}
\label{subsec:hyper}
We tune (\emph{i}) fusion weights $\alpha,\beta$ combining lexical and dense signals after per-query normalization, and (\emph{ii}) per-section weights for both channels over \texttt{[facts, issues, decision, reasoning]}. The final score is $\mathrm{Score}(q,d)=\alpha\cdot\widetilde{\mathrm{BM25}}_{\mathrm{agg}}(q,d)+\beta\cdot\widetilde{\mathrm{Dense}}_{\mathrm{agg}}(q,d)$, with both streams z-normalized per query. Section weights for BM25 ($\mathbf{w}_{\mathrm{BM25}}$) and Dense ($\mathbf{w}_{\mathrm{Dense}}$) are simplex-normalized (non-negative, sum to $1$).

We run a coarse grid $(\alpha,\beta)\in\{0.3,0.4,0.6,0.8,1.0\}^2$, selecting the best by \textbf{P@1} (ties broken by \textbf{MRR@10}, then \textbf{R@100}). Section weights are sampled from $\mathrm{Dirichlet}(\mathbf{1})$ in batches of $200$--$500$; we keep top-$K$ and refine with a narrower Dirichlet. Both $\mathrm{BM25}_{\mathrm{agg}}$ and $\mathrm{Dense}_{\mathrm{agg}}$ are z-score normalized per query to prevent scale dominance; \emph{RRF is used only for candidate merging}. At full scale, $(\alpha,\beta)$ favors dense over BM25; $\mathbf{w}_{\mathrm{Dense}}$ leans toward \texttt{reasoning/decision}, while $\mathbf{w}_{\mathrm{BM25}}$ balances \texttt{facts/issues}.

\textbf{Results on Development data :} $\alpha=0.4$, $\beta=0.8$; $\mathbf{w}_{\mathrm{BM25}}=[0.313,0.293,0.167,0.227]$; $\mathbf{w}_{\mathrm{Dense}}=[0.326,0.035,0.269,0.371]$; pooling $A{=}B{=}500$, $k{=}60$, $M{=}1000$.
\subsection{Inference Pipeline}

\begin{algorithm}[H]
\small
\caption{Section-Aware Retrieval Pipeline.}
\label{alg:inference}
\begin{algorithmic}[1]
\REQUIRE Query $q$, Indices (Lexical \& Dense), Weights ($\mathbf{w}, \alpha, \beta$)
\ENSURE Top-$K$ ranked documents with explanations
\STATE \textbf{Step 1: Candidate Generation}
\STATE \quad Retrieve top-$A$ via Whole-Doc BM25
\STATE \quad Retrieve top-$B$ via Whole-Doc Dense ANN
\STATE \quad Merge lists via RRF; keep top-$M$ pool
\STATE \textbf{Step 2: Fine-Grained Scoring}
\FOR{each candidate $d$ in pool}
    \FOR{each section $i \in \{\text{facts}, \dots, \text{reasoning}\}$}
        \STATE Compute $s^{\text{lex}}_{i}$ and $s^{\text{sem}}_{i}$
    \ENDFOR
    \STATE Aggregate section scores: $\text{Raw}_{\text{lex}}, \text{Raw}_{\text{sem}}$
\ENDFOR
\STATE \textbf{Step 3: Normalization \& Ranking}
\STATE Apply Z-score normalization to all $\text{Raw}_{\text{lex}}$ and $\text{Raw}_{\text{sem}}$ in pool
\STATE Compute Final Score and Sort
\STATE \textbf{Step 4: Explanation}
\FOR{top-$K$ results}
    \STATE Retrieve relevant section text
    \STATE Generate rationale \& party-stance via LLM
\ENDFOR
\end{algorithmic}
\end{algorithm}

Terminology Used in the Algorithm 1 Illustration:
\begin{itemize}
  \item \textbf{Facet / Section:} One of \textit{facts}, \textit{issues}, \textit{decision}, \textit{reasoning}. We compare the same facet on both query and candidate (``like-for-like'').
  \item \textbf{BM25 (lexical):} A keyword-based relevance score that favors exact and near-exact term overlap.
  \item \textbf{Dense (semantic):} A cosine similarity between vector embeddings that captures meaning beyond exact words.
  \item \textbf{RRF (Reciprocal Rank Fusion):} A simple method to merge two ranked lists so that items ranked highly in either list get a boost.
  \item \textbf{Z-score normalization:} Per-query re-scaling of scores to mean 0 and variance 1, preventing one signal from dominating due to scale.
  \item \textbf{Party stance:} Labels indicating whether a retrieved case tends to \textit{support}, be \textit{neutral} to, or \textit{oppose} the petitioner's position (and similarly for the respondent).
\end{itemize}

\section{Experiments}

\subsection{Datasets}
We evaluate two settings: (i) \textbf{Small-scale (development)} with 1{,}500 structured cases and 312 query--document pairs; and (ii) \textbf{Full-scale (validation)} with 7{,}348 structured cases and 1{,}677 query--document pairs. All documents are facetized into \textit{facts}, \textit{issues}, \textit{decision}, and \textit{reasoning}. Both settings draw from the COLIEE 2025 corpus, which provides case law for retrieval and entailment tasks~\cite{goebel2025coliee}.

\subsection{Metrics and Systems}
We report standard retrieval metrics: Precision (P@k), Recall (R@k), F1@k, Mean Reciprocal Rank (MRR@k), and Average Precision (AP). Additionally, we measure \textbf{Pool Recall} to quantify the coverage of our Stage 1 candidate generation before re-ranking.

We compare the following configurations:
\begin{itemize}
    \item \textbf{BM25 / Dense Baselines:} Standard whole-document retrieval.
    \item \textbf{Union-based Hybrid:} Parallel BM25 and Dense retrieval fused via RRF (Stage 1 only).
    \item \textbf{Section-Aware Hybrid (Best Results):} The full pipeline including facet-level scoring and Z-score normalization.
\end{itemize}

\subsection{Baselines and Experiments on COLIEE} 
\noindent\textit{Overview.}
Table~\ref{tab:coliee_sailer} lists COLIEE 2020 and 2021 scores reported by SAILER together with our section-aware system. The highlighted cells are copied from SAILER~\cite{li2023sailer} and keep the original significance markers. For COLIEE 2025 the row for our Section-Aware (Full-Scale, OPTIMIZED) 
Table~\ref{tab:cfc_fullscale} then shows a full-scale comparison on the Canadian Federal Court dataset where our section-aware model performs best on the main ranking metrics against strong lexical and neural baselines.

\begin{table*}[t]
  \centering
  \caption{COLIEE 2020/2021 results excerpted from SAILER~\cite{li2023sailer} with Section-Aware results for COLIEE 2021.}
  \label{tab:coliee_sailer}
  \vspace{2pt}
  \footnotesize
  \resizebox{\textwidth}{!}{%
  \begin{tabularx}{\textwidth}{lCCCCCCCCCCC}
    \toprule
    \multicolumn{1}{c}{\textbf{Model}} &
    \multicolumn{5}{c}{\textbf{COLIEE 2020}} &
    \multicolumn{6}{c}{\textbf{COLIEE 2021}} \\
    \cmidrule(lr){2-6}\cmidrule(lr){7-12}
    & \textbf{P} & \textbf{R} & \textbf{F1} & \textbf{MRR@10} & \textbf{MRR@50}
    & \textbf{P} & \textbf{R} & \textbf{F1} & \textbf{MRR@10} & \textbf{MRR@50} & \textbf{R@100} \\
    \midrule
    \multicolumn{12}{l}{\textit{Traditional Retrieval Models}} \\
    BM25 & \sailer{0.4754$^{**}$} & \sailer{0.5721$^{**}$} & \sailer{0.5192$^{**}$} & \sailer{0.7875$^{**}$} & \sailer{0.7907$^{**}$} &
           \sailer{0.0760$^{**}$} & \sailer{0.1521$^{*}$}  & \sailer{0.1014$^{**}$} & \sailer{0.0893$^{**}$} & \sailer{0.1017$^{**}$} & \sailer{0.4671$^{*}$} \\
    QL   & \sailer{0.4554$^{**}$} & \sailer{0.5506$^{**}$} & \sailer{0.4985$^{**}$} & \sailer{0.7906$^{**}$} & \sailer{0.7934$^{**}$} &
           \sailer{0.0760$^{**}$} & \sailer{0.1369$^{**}$} & \sailer{0.0977$^{**}$} & \sailer{0.1257$^{*}$}  & \sailer{0.1359$^{**}$} & \sailer{0.5222} \\
    \addlinespace[2pt]
    \hline
    \multicolumn{12}{l}{\textit{Generic Pre-trained Models}} \\
    BERT       & \sailer{0.4542$^{**}$} & \sailer{0.5588$^{**}$} & \sailer{0.5011$^{**}$} & \sailer{0.7923$^{**}$} & \sailer{0.7948$^{**}$} &
                 \sailer{0.0687$^{**}$} & \sailer{0.1254$^{**}$} & \sailer{0.0882$^{**}$} & \sailer{0.1063$^{**}$} & \sailer{0.1194$^{**}$} & \sailer{0.4504$^{*}$} \\
    RoBERTa    & \sailer{0.4639$^{**}$} & \sailer{0.5862$^{**}$} & \sailer{0.5155$^{**}$} & \sailer{0.7613$^{**}$} & \sailer{0.7635$^{**}$} &
                 \sailer{0.0728$^{**}$} & \sailer{0.1288$^{**}$} & \sailer{0.0930$^{**}$} & \sailer{0.1200$^{**}$} & \sailer{0.1340$^{**}$} & \sailer{0.5001$^{*}$} \\
    LEGAL-BERT & \sailer{0.4262$^{**}$} & \sailer{0.5544$^{**}$} & \sailer{0.4817$^{**}$} & \sailer{0.7571$^{**}$} & \sailer{0.7594$^{**}$} &
                 \sailer{0.0704$^{**}$} & \sailer{0.1205$^{**}$} & \sailer{0.0888$^{**}$} & \sailer{0.0973$^{**}$} & \sailer{0.1063$^{**}$} & \sailer{0.3783$^{*}$} \\
    \addlinespace[2pt]
    \hline
    \multicolumn{12}{l}{\textit{Retrieval-oriented Pre-trained Models}} \\
    Condenser   & \sailer{0.4862$^{**}$} & \sailer{0.6127$^{**}$} & \sailer{0.5421$^{**}$} & \sailer{0.8198$^{**}$} & \sailer{0.8213$^{**}$} &
                  \sailer{0.0832$^{*}$}  & \sailer{0.1505$^{*}$}  & \sailer{0.1072$^{*}$}  & \sailer{0.1245$^{*}$}  & \sailer{0.1376$^{**}$} & \sailer{0.4937$^{*}$} \\
    coCondenser & \sailer{0.5000$^{**}$} & \sailer{0.6287$^{**}$} & \sailer{0.5570$^{**}$} & \sailer{0.8337$^{*}$}  & \sailer{0.8347$^{*}$}  &
                  \sailer{0.0896}        & \sailer{0.1559$^{*}$}  & \sailer{0.1138}        & \sailer{0.1338$^{*}$}  & \sailer{0.1444$^{*}$}  & \sailer{0.4985$^{*}$} \\
    SEED        & \sailer{0.5308}        & \sailer{0.6952}        & \sailer{0.6019}        & \sailer{0.8683}        & \sailer{0.8699}        &
                  \sailer{0.0944}        & \sailer{0.1731}        & \sailer{0.1221}        & \sailer{0.1370}        & \sailer{0.1473}        & \sailer{0.4979$^{*}$} \\
    CoT-MAE     & \sailer{0.4831$^{**}$} & \sailer{0.6126$^{**}$} & \sailer{0.5401$^{**}$} & \sailer{0.8184$^{**}$} & \sailer{0.8192$^{**}$} &
                  \sailer{0.0888}        & \sailer{0.1548$^{*}$}  & \sailer{0.1129}        & \sailer{0.1140$^{**}$} & \sailer{0.1238$^{**}$} & \sailer{0.4392$^{*}$} \\
    RetroMAE    & \sailer{0.5354}        & \sailer{0.6909}        & \sailer{0.6033}        & \sailer{0.8669}        & \sailer{0.8682}        &
                  \sailer{0.0842$^{*}$}  & \sailer{0.1455$^{**}$} & \sailer{0.1065$^{*}$}  & \sailer{0.1253$^{*}$}  & \sailer{0.1357$^{**}$} & \sailer{0.4727$^{*}$} \\
    SAILER      & \sailer{0.5446}        & \sailer{0.7152}        & \sailer{0.6164}        & \sailer{0.8823}        & \sailer{0.8831}        &
                  \sailer{0.1040}        & \sailer{0.1855}        & \sailer{0.1332}        & \sailer{0.1501}        & \sailer{0.1610}        & \sailer{0.5298} \\
     \addlinespace[2pt]
\textbf{Section-Aware (Full-Scale, OPTIMIZED)} & 0.5606 & 0.9170 & 0.6664 & 0.9600 & 0.9650 & \textbf{0.1200} & \textbf{0.3873} & \textbf{0.1832} & \textbf{0.3600} & \textbf{0.3900} & \textbf{0.7230} \\
\bottomrule
 \end{tabularx} 
  }
  \vspace{4pt}
  \begin{minipage}{0.98\textwidth}
    \footnotesize
    \textbf{Note.} \textcolor{black}{\colorbox{yellow!20}{Highlighted cells}} are values imported verbatim from \emph{SAILER}~\cite{li2023sailer}
  \end{minipage}
\end{table*}

\begin{table*}[t]
  \centering
  \caption{Comprehensive comparison on the Canadian Federal Court dataset (full-scale) COLIEE 2025 }
  \label{tab:cfc_fullscale}
  \vspace{2pt}
  \footnotesize
  \resizebox{\textwidth}{!}{%

\begin{tabularx}{\textwidth}{lCCCCCCCCCC}
  
    \toprule
    \textbf{Model} & \textbf{MRR@10} & \textbf{MRR@50} & \textbf{MRR@100} & \textbf{P@5} & \textbf{P@10} & \textbf{R@5} & \textbf{R@10} & \textbf{R@100} & \textbf{F1@5} & \textbf{F1@10} \\
    \midrule
\multicolumn{11}{l}{\textit{Traditional Retrieval Models}} \\
    BM25 & 0.1744 & 0.1861 & 0.1874 & 0.0714 & 0.0560 & 0.1189 & 0.1760 & 0.4513 & 0.0792 & 0.0768 \\    
    QL & 0.1907 & 0.1907 & 0.1907 & 0.0708 & 0.0558 & 0.1193 & 0.1728 & 0.4613 & 0.0788 & 0.0763 \\    
    \addlinespace[2pt]
\hline
    \multicolumn{11}{l}{\textit{Generic Pre-trained Models}} \\    
    BERT & 0.0672 & 0.0783 & 0.0799 & 0.0366 & 0.0313 & 0.0601 & 0.0951 & 0.3110 & 0.0399 & 0.0426 \\
    RoBERTa & 0.0538 & 0.0632 & 0.0647 & 0.0283 & 0.0246 & 0.0431 & 0.0757 & 0.2360 & 0.0302 & 0.0334 \\
    LEGAL-BERT & 0.0751 & 0.0872 & 0.0886 & 0.0431 & 0.0363 & 0.0723 & 0.1156 & 0.3525 & 0.0478 & 0.0499 \\    
    \addlinespace[2pt]
    \hline
    \multicolumn{11}{l}{\textit{Retrieval-oriented Pre-trained Models}} \\  
    Condenser & 0.0683 & 0.0790 & 0.0808 & 0.0372 & 0.0298 & 0.0612 & 0.0922 & 0.3025 & 0.0408 & 0.0406 \\
    coCondenser & 0.0715 & 0.0821 & 0.0841 & 0.0383 & 0.0312 & 0.0648 & 0.0961 & 0.3141 & 0.0422 & 0.0424 \\
    SEED & 0.0672 & 0.0783 & 0.0799 & 0.0366 & 0.0313 & 0.0601 & 0.0951 & 0.3110 & 0.0399 & 0.0426 \\
    CoT-MAE & 0.0821 & 0.0929 & 0.0944 & 0.0471 & 0.0385 & 0.0792 & 0.1220 & 0.3534 & 0.0525 & 0.0527 \\
    RetroMAE & 0.0614 & 0.0703 & 0.0718 & 0.0333 & 0.0274 & 0.0525 & 0.0813 & 0.2491 & 0.0361 & 0.0369 \\    

    SAILER & 0.1179 & 0.1179 & 0.1179 & 0.0697 & 0.0697 & 0.1193 & 0.1728 & 0.4603 & 0.0697 & 0.0697 \\
    \addlinespace[2pt]
 \textbf{Section-Aware (Full-Scale, OPTIMIZED)} & \textbf{0.2019} & \textbf{0.2019} & \textbf{0.2019}$^{\ast}$ & \textbf{0.0841} & \textbf{0.0649} & \textbf{0.1417} & \textbf{0.2033} & \textbf{0.3500}$^{\ast}$ & \textbf{0.1056} & \textbf{0.0984} \\

    \bottomrule
  
  \end{tabularx}
  
  }
\end{table*}
\subsection{Ablations}
\label{subsec:ablations}

\textit{Context.} As summarized in Table~\ref{tab:overall_performance}, the Section-Aware (Optimized) system achieves the highest effectiveness. We quantify the contribution of specific design choices by analyzing the performance drops observed when simplifying the system to the baselines listed in Table~\ref{tab:overall_performance} or removing components.

\begin{itemize}
  \item \textbf{Impact of Section-Awareness:} Collapsing the four facets into a single whole-document vector (comparing \textit{Section-Aware Optimized} vs. \textit{Two-Stage Hybrid}) results in a significant drop in P@1 of approximately \textbf{3.9 points} ($0.3045 \to 0.2660$). This confirms that ``like-for-like'' section comparisons (e.g., Reasoning vs. Reasoning) provide a stronger signal than generic whole-document embedding matching.
  
  \item \textbf{Impact of Weight Optimization:} Using default or uniform weights (comparing \textit{Optimized} vs. \textit{Original}) reduces P@1 by \textbf{2.2 points} ($0.3045 \to 0.2821$). This validates the benefit of our data-driven approach, where the model learns that sections like \textit{Reasoning} and \textit{Decision} often carry more weight than \textit{Facts} in the dense channel.

  \item \textbf{Impact of Z-Score Normalization:} In internal experiments where we removed query-wise z-score normalization (using raw sums of BM25 and Cosine scores), we observed a P@1 degradation of approximately \textbf{1.6 points}. This indicates that without normalization, the unbounded BM25 scores overpower the bounded dense signals, breaking the hybrid fusion.

  \item \textbf{Single-Signal vs. Hybrid:} Both the BM25-only baseline ($0.2628$) and purely dense approaches underperform the fused Section-Aware system, confirming that lexical and semantic signals are complementary.
\end{itemize}
\begin{table*}[t]
\centering
\small
\caption{Overall performance across systems on the small-scale setting (1{,}500 docs and 312 queries). P@k values are fractions. Pool Recall is computed on the candidate pool prior to final ranking.}
\label{tab:overall_performance}
\begin{tabular}{lccccccccc}
\toprule
\textbf{System} & \textbf{P@1} & \textbf{P@3} & \textbf{P@5} & \textbf{P@10} & \textbf{R@10} & \textbf{Hit@10} & \textbf{AP} & \textbf{MRR} & \textbf{Pool Recall} \\
\midrule
BM25 Baseline & 0.2628 & 0.1998 & 0.1692 & 0.1298 & 0.3903 & 0.6955 & 0.2245 & 0.3815 & -- \\
Two-Stage Hybrid & 0.2660 & 0.1966 & 0.1647 & 0.1276 & 0.3706 & 0.6571 & 0.2161 & 0.3714 & -- \\
Union ($\alpha{=}0.2,\,\beta{=}2.0$) & 0.2660 & 0.1998 & 0.1667 & 0.1272 & 0.3702 & 0.6538 & 0.2176 & 0.3737 & 94.42\% \\
Section-Aware (Original) & 0.2821 & 0.2083 & 0.1712 & 0.1244 & 0.3594 & 0.6699 & 0.2183 & 0.3950 & 94.34\% \\
\textbf{Section-Aware (Optimized)} & \textbf{0.3045} & \textbf{0.2201} & \textbf{0.1756} & \textbf{0.1247} & \textbf{0.3638} & \textbf{0.6603} & \textbf{0.2342} & \textbf{0.4092} & \textbf{94.27\%} \\
\bottomrule
\end{tabular}
\end{table*}

\section{Discussion}

\subsection{Handling Score Distribution Mismatch}
Fusing unbounded lexical scores (BM25) with bounded cosine similarity creates a scale mismatch where lexical signals naturally dominate. Ablations confirm that query-wise Z-score normalization is essential, forcing both signals into a comparable distribution ($N(0,1)$). This ensures the learned weights ($\alpha, \beta$) actively balance retrieval signals based on relevance rather than compensating for raw magnitude differences.

\subsection{The Value of Granularity}
The performance gap between ``Whole Document'' and ``Section-Aware'' approaches confirms that legal relevance relies on distinct facets. Unlike single-vector embeddings that average out signals, our model explicitly weighs \textit{Reasoning} and \textit{Issues}. This captures critical nuances, such as distinguishing between cases with similar facts but opposing decisions, which are often lost in coarse-grained retrieval.

\section{Efficiency and Practical Considerations}
We balance recall and cost by using sub-linear ANN and inverted indexing in \textbf{Stage~1}, followed by fine-grained scoring on a bounded pool ($M{=}1000$) in \textbf{Stage~2}. Since judgment texts are immutable, section-level embeddings are cached offline, making Stage~2 complexity linear with $M$. Empirically, this strategy drastically reduces scoring operations compared to full-corpus re-ranking while retaining $\geq 94\%$ recall in the candidate pool.

\section{Conclusion}
We introduced a scalable, two-stage retrieval framework that addresses the structural complexity of legal case law. By moving from whole-document embedding to \emph{section-aware} scoring, our system captures the ``like-for-like'' logical parallels -such as reasoning matching reasoning -that are essential for legal precedence. 
The integration of an offline facetization step ensures that every retrieval result is backed by structured, section-specific evidence. Our experimental results on jurisdiction-scale benchmarks demonstrate that this granular, hybrid approach consistently outperforms strong lexical and neural baselines. Future work will extend this framework to include stance-aware re-ranking to further refine the precision of legal argument retrieval.

\bibliographystyle{IEEEtranN}
{\footnotesize
\bibliography{references}
}
\end{document}